\RequirePackage{ifpdf}
\ifpdf % We are running pdfTeX in pdf mode
\documentclass[pdftex]{sigma}
\else
\documentclass{sigma}
\fi

\numberwithin{equation}{section}

\begin{document}

\allowdisplaybreaks

\renewcommand{\PaperNumber}{024}

\FirstPageHeading

\ShortArticleName{B\"acklund Transformations  for First and Second
Painlev\'{e} Hierarchies}

\ArticleName{B\"acklund Transformations\\ for First and Second
Painlev\'{e} Hierarchies}

\Author{Ayman Hashem SAKKA}

\AuthorNameForHeading{A.H. Sakka}

\Address{Department of Mathematics,
             Islamic University of Gaza, P.O.\ Box 108, Rimal, Gaza, Palestine}
\Email{\href{mailto:asakka@iugaza.edu.ps}{asakka@iugaza.edu.ps}}

\ArticleDates{Received November 25, 2008, in f\/inal form February 24,
2009; Published online March 02, 2009}

\Abstract{We give B\"acklund transformations for f\/irst and second
Painlev\'{e} hierarchies. These B\"acklund transformations are
generalization of known B\"acklund transformations of the f\/irst
and second Painlev\'{e} equations and they relate the considered
hierarchies to new hierarchies of Painlev\'{e}-type equations.}

\Keywords{Painlev\'{e} hierarchies; B\"acklund transformations}

\Classification{34M55; 33E17}

\section{Introduction}
One century  ago Painlev\'{e} and Gambier have discovered the six
Painlev\'{e} equations, PI--PVI. These equations are  the only
second-order ordinary dif\/ferential equations whose general
solutions can not be expressed in terms of elementary and
classical special functions; thus they def\/ine new transcendental
functions.
 Painlev\'{e} transcendental functions appear in many areas of
 modern mathematics and physics and they paly the same role in
 nonlinear problems as the classical special functions play in
 linear problems.

In recent years there is a considerable interest in studying
hierarchies of Painlev\'{e} equations.
This interest is due to the connection between these hierarchies
of Painlev\'{e} equations and completely integrable partial
dif\/ferential equations.
A Painlev\'{e} hierarchy is an inf\/inite sequence of nonlinear
ordinary dif\/ferential equations whose f\/irst member is a
Painlev\'{e} equation.
  Airault~\cite{Airault} was the f\/irst to derive a Painlev\'{e} hierarchy,
namely   a second Painlev\'{e} hierarchy, as the similari\-ty
reduction  of the modif\/ied Korteweg--de Vries (mKdV) hierarchy. A
f\/irst Painlev\'{e} hierarchy was given by Kudryashov~\cite{Kudr3}.
Later on  several hierarchies of Painlev\'{e} equations were
introdu\-ced~\cite{Hone,Kud2,Kud,Kudr4,R5,R5.1,R6,GJP1,AS4}.

As it is well known, Painlev\'{e} equations possess B\"acklund
transformations; that is,  mappings between solutions of the same
Painlev\'{e} equation or between solutions of  a particular
Painlev\'{e} equation  and other second-order Painlev\'{e}-type
equations. Various methods to derive these  B\"acklund
transformations can be found  for example in
\cite{Fokas,Okamoto3,Gordoa,Gromak}. B\"acklund transformations are
nowadays considered to be one of the main properties  of
integrable nonlinear ordinary dif\/ferential equations, and there is
much interest in their derivation.

In the present article, we  generalize known B\"acklund
transformations of the f\/irst and se\-cond Painlev\'{e} equations to
the f\/irst and se\-cond Painlev\'{e} hierarchies given in
\cite{Kudr4, AS4}. We give a~B\"acklund transformation between the
considered f\/irst Painlev\'{e} hierarchy and a new hierarchy of
Painlev\'{e}-type equations. In addition, we  give two new
hierarchies of Painlev\'{e}-type equations related, via B\"acklund
transformations, to the considered second Painlev\'{e} hierarchy.
Then we derive auto-B\"acklund transformations for this second
Painlev\'{e} hierarchy. B\"acklund transformations of the second
Painlev\'{e} hierarchy have been studied in \cite{Kudr4, R4}.

\section{B\"acklund transformations for PI hierarchy}

In this section, we will derive a B\"acklund transformation for
the f\/irst Painlev\'{e} hierarchy (PI hierarchy) \cite{Kudr4}
\begin{gather}\label{K-PI-h}
 \sum_{j=2}^{n+1}\gamma_jL^{j}[u]=\gamma x,
\end{gather}
where
 the operator $L^j[u]$   satisf\/ies  the Lenard recursion
relation
\begin{gather}\label{LR}
    D_xL^{j+1}[u]=\big(D_x^3-4uD_x-2u_x\big)L^j[u], \qquad L^1[u]=u.
\end{gather}
The special case $\gamma_j=0$, $2\leq j\leq n$, of this  hierarchy
is a similarity reduction of the Schwarz--Korteweg--de Vries
hierarchy \cite{Kudr3, Kud2}. Moreover its members may def\/ine new
transcendental functions.

The PI hierarchy (\ref{K-PI-h}) can be written in the following
form  \cite{AS4}
\begin{gather}\label{PI-h}
 \mathcal{R}_{_{\rm I}}^nu+\sum_{j=2}^{n}\kappa_j\mathcal{R}_{_{\rm I}}^{n-j}u=x,
\end{gather}
where $\mathcal{R}_{_{\rm I}}$ is the recursion operator
\[
    \mathcal{R}_{_{\rm I}}=D_x^2-8u+4D_x^{-1}u_x.
\]

In \cite{Cos, AS3}, it is shown that the  B\"acklund
transformation
\begin{gather}\label{PI-BT-m=1}
  u=-y_x,\qquad y=\tfrac{1}{2}\big(u_x^2-4u^3-2xu\big),
\end{gather}
 def\/ines  a one-to-one correspondence between
 the f\/irst Painlev\'{e} equation
\begin{gather}\label{PI}
    u_{xx}=6u^2+x.
\end{gather}
 and the  SD-I.e equation of Cosgrove and
Scouf\/is \cite{Cos}
\begin{gather}\label{SD-I.e}
    y_{xx}^2=-4y_x^3-2(xy_x-y).
\end{gather}

We will show that there is a generalization of this  B\"acklund
transformation to all members of the PI hierarchy (\ref{PI-h}).
Let
\begin{gather}\label{PI-BT-1}
y=-xu+D_x^{-1}u_x\Bigg[\mathcal{R}_{_{\rm I}}^nu+\sum_{j=2}^{n}\kappa_j\mathcal{R}_{_{\rm I}}^{n-j}u\Bigg].
\end{gather}
Dif\/ferentiating (\ref{PI-BT-1}) and using (\ref{PI-h}), we f\/ind
\begin{gather}\label{PI-BT-2}
    u=-y_x.
\end{gather}
Substituting $u=-y_x$ into  (\ref{PI-BT-1}), we obtain  the
following hierarchy of dif\/ferential equation for $y$
\begin{gather}\label{SD-I.e-h}
D_x^{-1}y_{xx}\Bigg[\mathcal{S}_{_{\rm I}}^ny_x+\sum_{j=2}^{n}\kappa_j\mathcal{S}_{_{\rm I}}^{n-j}y_x\Bigg]+(xy_x-y)=0,
\end{gather}
where $\mathcal{S}_{_{\rm I}}$ is the recursion operator
\[
    \mathcal{S}_{_{\rm I}}=D_x^2+8y_x-4D_x^{-1}y_{xx}.
\]
The f\/irst member of the hierarchy (\ref{SD-I.e-h}) is the SD-I.e
equation (\ref{SD-I.e}). Thus we shall call this hierarchy SD-I.e
hierarchy.

Therefor we have derived the B\"acklund transformation
(\ref{PI-BT-1})--(\ref{PI-BT-2}) between solutions $u$ of the f\/irst
Painlev\'{e} hierarchy (\ref{PI-h}) and solutions $y$ of the
SD-I.e hierarchy (\ref{SD-I.e-h}).

When $n=1$, the B\"acklund transformation
(\ref{PI-BT-1})--(\ref{PI-BT-2}) gives the B\"acklund transformation~(\ref{PI-BT-m=1}) between the f\/irst Painlev\'{e} equation
(\ref{PI}) and the SD-I.e equation (\ref{SD-I.e}). Next we will
consider the cases $n=2$ and $n=3$.
%===============================================================
%            Example (1)
%===============================================================

\begin{example}[$n=2$]
%{\bf Example (1):}
The second member of the PI hierarchy (\ref{PI-h}) is the
fourth-order equation
\begin{gather}\label{PI-O4}
     u_{xxxx}=20uu_{xx}+10u_x^2-40u^3-\kappa_2u+x.
\end{gather}
In this case,  the B\"acklund transformation (\ref{PI-BT-1}) reads
\begin{gather}\label{PI-O4-BT}
    y=\tfrac{1}{2}\big(2u_xu_{xxx}-u_{xx}^2-20uu_x^2+20u^4+\kappa_2u^2-2xu\big).
\end{gather}
Equations (\ref{PI-O4-BT}) and (\ref{PI-BT-2}) give one-to-one
correspondence between (\ref{PI-O4}) and   the following equation
\begin{gather}\label{PI-O4-e}
    2y_{xx}y_{xxxx}-y_{xxx}^2+20y_xy_{xx}^2+20y_x^4+\kappa_2y_x^2+2(xy_x-y)=0.
\end{gather}
Equation (\ref{PI-O4-e}) and the  B\"acklund transformation
(\ref{PI-BT-2}) and  (\ref{PI-O4-BT}) were given before~\cite{AS1}.
\end{example}

%===============================================================
%            Example (3)
%===============================================================

\begin{example}[$n=3$]
%{\bf Example (2):}
The third member of the PI hierarchy (\ref{PI-h}) reads
\begin{gather}
    u_{xxxxxx}=28uu_{xxxx}+56u_xu_{xxx}+42u_{xx}^2-280u^2u_{xx}\nonumber\\
\phantom{u_{xxxxxx}=}{}  -280uu_x^2+280u^4-\kappa_2\big(u_{xx}-6u^2\big)-\kappa_3u+x.\label{PI-O6}
\end{gather}
In this case, the B\"acklund transformation (\ref{PI-BT-1}) has
the form
\begin{gather}
    y=\tfrac{1}{2}\big[2u_xu_{xxxxx}-2u_{xx}u_{xxxx}+u_{xxx}^2-56uu_xu_{xxx}+28uu_{xx}^2\nonumber\\
\phantom{y=}{}
-56u_x^2u_{xx}+280u^2u_x^2-112u^5+\kappa_2\big(u_x^2-4u^3\big)+\kappa_3u^2-2xu\big].\label{PI-O6-BT1}
\end{gather}
Equations (\ref{PI-BT-2}) and (\ref{PI-O6-BT1})   give one-to-one
correspondence between solutions $u$ of (\ref{PI-O6}) and
solutions $y$ of the following equation
\begin{gather}
    2y_{xx}y_{xxxxxx}-2y_{xxx}y_{xxxxx}+y_{xxxx}^2+56y_xy_{xx}y_{xxxx}-28y_xy_{xxx}^2\nonumber\\
\qquad{}+56y_{xx}^2y_{xxx}+280y_x^2y_{xxx}^2+112y_x^5+\kappa_2\big(y_{xx}^2+4y_x^3\big)+\kappa_3y_x^2+2(xy_x-y)=0.\label{PI-O6-y}
\end{gather}
Equation (\ref{PI-O6-y}) is a new sixth-order Painlev\'{e}-type
equation.
\end{example}

\section{B\"acklund transformations for second Painlev\'{e}  hierarchy}

In the present section, we will study B\"acklund transformations
of  the second Painlev\'{e} hierarchy (PII hierarchy)~\cite{Kudr4}
\[
(D_x-2u)\sum_{j=1}^{n}\gamma_{j}L^{j}\big[u_x+u^2\big]+2\gamma
xu-\gamma-4\delta=0,
\]
where
 the operator $L^j[u]$ is def\/ined by (\ref{LR}).
The special case $\gamma_j=0, ~1\leq j\leq n-1,$ of this hierarchy
is a similarity reduction of the modif\/ied Korteweg--de Vries
hierarchy \cite{Kudr3, Kud2}. The members of this hierarchy may
def\/ine new transcendental functions.

 This hierarchy can be written in the following alternative form  \cite{AS4}
\begin{gather}\label{PII-h}
\mathcal{R}_{_{\rm II}}^nu+\sum_{j=1}^{n-1}\kappa_{j}\mathcal{R}_{_{\rm II}}^{j}u-(xu+\alpha)=0,
\end{gather}
where $\mathcal{R}_{_{\rm II}}$ is the recursion operator
\[
 \mathcal{R}_{_{\rm II}}=D^2_x-4u^2+4uD^{-1}_xu_x.
\]

\subsection{A hierarchy of    SD-I.d equation}

As a f\/irst B\"acklund transformation for the PII hierarchy
(\ref{PII-h}), we will generalize the B\"acklund transformation
between the second Painlev\'{e} equation and the SD-I.d equation
of Cosgrove and Scouf\/is~\cite{Cos, AS3}.

Let
\begin{gather}\label{PII-BT-1}
y=D_x^{-1}\Bigg[u_x\Bigg(\mathcal{R}_{_{\rm II}}^nu+\sum_{j=1}^{n-1}\kappa_{j}\mathcal{R}_{_{\rm II}}^{j}u\Bigg)\Bigg]
-\tfrac{1}{2}xu^2-\tfrac{1}{2}(2\alpha-\epsilon)u,
\end{gather}
where $\epsilon=\pm 1$.  Dif\/ferentiating (\ref{PII-BT-1}) and
using (\ref{PII-h}), we f\/ind
\begin{gather}\label{PII-BT-2}
    u_x=\epsilon\big(u^2+2y_x\big).
\end{gather}

Now we will show that
\begin{gather}\label{PII-p5}
    D_x^{-1}\big(u_x
    \mathcal{R}_{_{\rm II}}^ju\big)=\tfrac{1}{2}\big(u^2H^j[y_x]+D_x^{-1}y_xH^j_x[y_x]\big),
\end{gather}
where the operator $H^j[p]$  satisf\/ies  the Lenard recursion
relation
\begin{gather}\label{LR2}
    D_xH^{j+1}[p]=\big(D_x^3+8pD_x+4p_x\big)H^j[p], \qquad H^1[p]=4p.
\end{gather}

Firstly, we will use induction to show that for any $j=1,2,\dots$,
\begin{gather}\label{PII-p1}
 \mathcal{R}_{_{\rm II}}^ju=\tfrac{1}{2}(\epsilon D_x+2u)H^j[y_x].
\end{gather}

For $j=1$,  $\mathcal{R}_{_{\rm II}}u=u_{xx}-2u^3$. Using
(\ref{PII-BT-2}), we f\/ind that
\begin{gather}\label{PII-p11}
u_{xx}=2u^3+4y_xu+2\epsilon y_{xx}.
\end{gather}
Thus
\begin{gather*}
\mathcal{R}_{_{\rm II}}u=4uy_x+2\epsilon y_{xx}  =\tfrac{1}{2}(\epsilon D_x+2u)H^1[y_x].
\end{gather*}

 Assume that it is true for $j=k$. Then
\begin{gather}
       2\mathcal{R}_{_{\rm II}}^{k+1}u=\mathcal{R}_{_{\rm II}}(\epsilon D_x+2u)H^k[y_x]
  =\epsilon H_{xxx}^k[y_x]+2uH^k_{xx}[y_x]+4u_xH^k_x[y_x]+2u_{xx}H^k[y_x]\nonumber\\
\phantom{2\mathcal{R}_{_{\rm II}}^{k+1}u=}{} -4u^2\big(\epsilon H^k_x[y_x]+2uH^k[y_x]\big)
       +4uD_x^{-1}\big(\epsilon u_xH^k_x[y_x]+2uu_xH^k[y_x]\big).\label{PII-p2}
\end{gather}
Integration  by parts gives
\[
D_x^{-1}\big(\epsilon
u_xH_x^k[y_x]+2uu_xH^k[y_x]\big)=u^2H^k[y_x]+D_x^{-1}\big[\big(\epsilon
u_x-u^2\big)H^k_x[y_x]\big].
\]
Hence (\ref{PII-p2}) can be written as
\begin{gather}
           2\mathcal{R}_{_{\rm II}}^{k+1}u
      =\epsilon H_{xxx}^k[y_x]+2uH^k_{xx}[y_x]+4u_xH^k_x[y_x]+2u_{xx}H^k[y_x]
  -4u^2\big(\epsilon H^k_x[y_x]+2uH^k[y_x]\big)\nonumber\\
\phantom{2\mathcal{R}_{_{\rm II}}^{k+1}u=}{} +4u\big(u^2H^k[y_x]+D_x^{-1}\big[\big(\epsilon
u_x-u^2\big)H^k_x[y_x]\big] \big).\label{PII-p14}
\end{gather}
 Using (\ref{PII-BT-2}) to substitute
$u_x$ and (\ref{PII-p11}) to substitute $u_{xx}$, (\ref{PII-p14})
becomes
\begin{gather*}
          2\mathcal{R}_{_{\rm II}}^{k+1}u=\epsilon
      \big(H_{xxx}^k[y_x]+8y_xH^k_x[y_x]+4y_{xx}H^k[y_x]\big)\nonumber\\
\phantom{2\mathcal{R}_{_{\rm II}}^{k+1}u=} {}+2u
      \big(H_{xx}^k[y_x]+4y_xH^k[y_x]+4D_x^{-1}y_{x}H_x^k[y_x]\big)\nonumber\\
\phantom{2\mathcal{R}_{_{\rm II}}^{k+1}u} {}      =(\epsilon
      D_x+2u)\big(H_{xx}^k[y_x]+4y_xH^k[y_x]+4D_x^{-1}y_{x}H_x^k[y_x]\big).\nonumber
\end{gather*}
Since
\[
      D_x\big(H_{xx}^k[y_x]+4y_xH^k[y_x]+4D_x^{-1}y_{x}H_x^k[y_x]\big)=H^k_{xxx}[y_x]+8y_xH^k_x[y_x]+4y_{xx}H^k[y_x],
\]
we have
$H_{xx}^k[y_x]+4y_xH^k[y_x]+4D_x^{-1}y_{x}H_x^k[y_x]=H^{k+1}[y_x]$,
see (\ref{LR2}),  and hence the proof by induction is f\/inished.

Now using (\ref{PII-p1}) we f\/ind
\begin{gather}\label{PII-p6}
    2u_x \mathcal{R}_{_{\rm II}}^k(u)=\big(\epsilon u_x-u^2\big)H_x^k[y_x]+D_x\big(u^2H^k[y_x]\big).
\end{gather}
Using (\ref{PII-BT-2}) to substitute $u_x$ into (\ref{PII-p6}) and
then  integrating, we obtain (\ref{PII-p5}).

  Therefore (\ref{PII-BT-1}) can be used to
obtain the following quadratic equation for $u$
\begin{gather}
 \Bigg(-x+H^n[y_x]+\sum_{j=1}^{n-1}\kappa_jH^j[y_x]\Bigg)u^2
    - (2\alpha-\epsilon )u\nonumber\\
     \qquad {}+ 2D_x^{-1}y_{x}\Bigg(H_x^n[y_x]+\sum_{j=1}^{n-1}\kappa_jH_x^j[y_x]\Bigg)-2y=0 .\label{PII-QE}
\end{gather}
Eliminating $u$ between  (\ref{PII-BT-2}) and (\ref{PII-QE}) gives
a one-to-one correspondence between the second Painlev\'{e}
hierarchy (\ref{PII-h}) and the following hierarchy of
second-degree equations
\begin{gather}
\Bigg(H_x^n[y_x]+\sum_{j=1}^{n-1}\kappa_jH_x^j[y_x]-1\Bigg)^2+8 \Bigg(H^n[y_x]+\sum_{j=1}^{n-1}\kappa_jH^j[y_x]-x\Bigg)\nonumber\\
\qquad
{}\times
\Bigg(D_x^{-1}y_xH^n_x[y_x]+\sum_{j=1}^{n-1}\kappa_jD_x^{-1}y_xH_x^j[y_x]-y\Bigg)  =(2\alpha-\epsilon)^2.\label{SD-I.d-h}
\end{gather}

Therefore we have derived the B\"{a}cklund transformation
(\ref{PII-BT-1}) and (\ref{PII-QE}) between the PII hierarchy
(\ref{PII-h}) and the new hierarchy (\ref{SD-I.d-h}).

Next we will give the explicit forms of the above results when
$n=1,2,3$.

\begin{example}[$n=1$]
The f\/irst member of the second Painlev\'{e} hierarchy
(\ref{PII-h})
 is the second Painlev\'{e} equation
 \[
    u_{xx}=2u^3+xu+\alpha.
\]
 In this case, (\ref{PII-BT-1}) and (\ref{PII-QE}) read
\[
    y=\tfrac{1}{2}\big[u_x^2-u^4-xu^2-(2\alpha-\epsilon)u\big]
\]
and
\[
    (4y_x-x)u^2-(2\alpha-\epsilon)u+4y_x^2-2y=0,
\]
respectively. The second-degree equation for $y$ is
\begin{gather}\label{PII-O2-SD}
    (4y_{xx}-1)^2+8(4y_x-x)\big(2y_x^2-y\big)=(2\alpha-\epsilon)^2.
\end{gather}
The change of variables $w=y-\frac{1}{8}x^2$ transforms
(\ref{PII-O2-SD}) into the SD-I.d equation of Cosgrove and Scouf\/is
\cite{Cos}
\[
    w_{xx}^2+4w_x^3+2w_x(xw_x-w)=\tfrac{1}{16}(2\alpha-\epsilon)^2.
\]

Thus when $n=1$, the B\"acklund transformation (\ref{PII-BT-1})
and (\ref{PII-QE}) is the known B\"acklund transformation between
the second Painlev\'{e} equation and  the SD-I.d equation
(\ref{SD-I.d-h}). Since the f\/irst member of the hierarchy
(\ref{SD-I.d-h}) is the SD-I.d equation, we shall call it SD-I.d
hierarchy.
\end{example}

\begin{example}[$n=2$]
The second member of the second Painlev\'{e} hierarchy
(\ref{PII-h}) reads
\begin{gather}\label{PII-O4}
    u_{xxxx}=10u^2u_{xx}+10uu_x^2-6u^5-\kappa_1\big(u_{xx}-2u^3\big)+xu+\alpha.
\end{gather}
Equation (\ref{PII-O4}) is labelled in \cite{R1, R2} as F-XVII.

In this case, (\ref{PII-BT-1}) and (\ref{PII-QE}) read
\begin{gather}\label{PII-O4-BT1}
    y=\tfrac{1}{2}\big[2u_xu_{xxx}-u_{xx}^2-10u^2u_x^2+2u^6+\kappa_1\big(u_x^2-u^4\big)-xu^2-(2\alpha-\epsilon)u\big]
\end{gather}
and
\begin{gather}
    \big(4y_{xxx}\!+24y_x^2\!+4\kappa_1y_x-x\big)u^2\!-(2\alpha-\epsilon)u +8y_xy_{xxx}\!-4y_{xx}^2\!+32y_x^3\!+4\kappa_1y_x^2\!-2y=0,\!\!\!\label{PII-O4-BT2}
\end{gather}
respectively. Equations (\ref{PII-O4-BT1}) and  (\ref{PII-O4-BT2})
give one-to-one correspondence between (\ref{PII-O4}) and the
following fourth-order second-degree equation
\begin{gather}
    [4y_{xxxx}+48y_xy_{xx}+4\kappa_1y_{xx}-1]^2   \label{PII-O4-SD}   \\
   \qquad{} +8\big[4y_{xxx}+24y_x^2+4\kappa_1y_x-x\big]
    \big[4y_xy_{xxx}-2y_{xx}^2+16y_x^3+2\kappa_1y_x^2-y\big]=(2\alpha-\epsilon)^2. \nonumber
 \end{gather}
 Equation (\ref{PII-O4-SD}) is a f\/irst integral of the following
 f\/ifth-order equation
 \begin{gather}\label{Fif-1}
    y_{xxxxx}=-20y_xy_{xxx}-10y_{xx}^2-40y_x^3-\kappa_1y_{xxx}-6\kappa_1y_x^2+xy_x+y.
\end{gather}
The transformation $y=-(w+\frac{1}{2}\gamma
z+5\gamma^3)$, $z=x+30\gamma^2$ transforms (\ref{Fif-1}) into the
equation
\begin{gather}\label{Fif-2}
 w_{zzzzz}=20w_zw_{zzz}+10w_{zz}^2-40w_z^3+zw_z+w+\gamma z.
\end{gather}
The B\"acklund transformation \cite{ AS11}
\begin{gather}
    v=w_z,\qquad
    w=v_{zzzz}-20vv_{zz}-10v_z^2+40v^3-zv-\gamma z,
\end{gather}
gives a one-to-one correspondence between (\ref{Fif-2}) and
Cosgrove's Fif-III equation \cite{R1}
\begin{gather}\label{Fif-III}
 v_{zzzzz}=20vv_{zzz}+40v_zv_{zz}-120v^2v_z+zv_z+2v+\gamma.
\end{gather}
Therefore we have rederived the known relation
\[
    v=-\frac{1}{2}\big(\epsilon u_x-u^2+\gamma\big),\qquad
    u=\frac{-\epsilon[v_{zzz}-12vv_z+4\gamma v_z+\frac{\epsilon}{2}\alpha]}{2[v_{zz}-6v^2+4\gamma
    v+\frac{1}{4}z-4\gamma^2]}.
\]
between Cosgrove's equations Fif-III (\ref{Fif-III}) and F-XVII
(\ref{PII-O4}) \cite{R1}.
\end{example}

\begin{example}[$n=3$]
The third member of the second Painlev\'{e} hierarchy
(\ref{PII-h}) reads
\begin{gather}
    u_{xxxxxx}=14u^2u_{xxxx}+56uu_xu_{xxx}+42uu_{xx}^2+70u_x^2u_{xx}-70u^4u_{xx}-140u^3u_x^2+20u^7\nonumber\\
   \phantom{u_{xxxxxx}=}{}
-\kappa_2(u_{xxxx}-10u^2u_{xx}-10uu_x^2+6u^5)
  -\kappa_1(u_{xx}-2u^3)+xu+\alpha.\label{PII-O6}
\end{gather}
In this case, (\ref{PII-BT-1}) and (\ref{PII-QE}) have the
following forms respectively
\begin{gather}
    2y=2u_xu_{xxxxx}-2u_{xx}u_{xxxx}+u^2_{xxx}-28u^2u_xu_{xxx}
     +14u^2u_{xx}^2\!-56uu_x^2u_{xx}-21u_x^4\!+70u^4u_x^2\!\nonumber\\
     \phantom{2y=}{}-5u^8
   +\kappa_2(2u_xu_{xxx}-u_{xx}^2-10u^2u_x^2+2u^6)
     +\kappa_1(u_x^2-u^4) -xu^2-(2\alpha-\epsilon)u \label{PII-O6-BT1}
\end{gather}
and
\begin{gather}
    4\left[y_{xxxxx}+20y_xy_{xx}+10y_{xx}^2+40y_x^3+\kappa_2\big(y_{xxx}+6y_x^2\big)+\kappa_1y_x-\tfrac{1}{4}x\right]u^2
    \nonumber\\
    \qquad{}-(2\alpha-\epsilon)u+4\big(2y_xy_{xxxxx}-2y_{xx}y_{xxxx}+y_{xxx}^2+40y_x^2y_{xxx}+60y_x^4\big)\nonumber\\
    \qquad{}+4\kappa_2\big(2y_xy_{xxx}-y_{xx}^2+8y_x^3\big)+4\kappa_1y_x^2-2y=0. \label{PII-O6-BT2}
\end{gather}
Equations (\ref{PII-O6-BT1}) and (\ref{PII-O6-BT2}) give
one-to-one correspondence between (\ref{PII-O6}) and the following
six-order second-degree equation
\begin{gather}
    \left[y_{xxxxxx}+20y_xy_{xxxx}+40y_{xx}y_{xxx}
    +120y_x^2y_{xx}
    +\kappa_2(y_{xxxx}+12y_xy_{xx})+\kappa_1y_{xx}-\tfrac{1}{4}\right]^2 \nonumber\\
    \qquad {}+2\left[y_{xxxxx}+20y_xy_{xx}+10y_{xx}^2+40y_x^3+\kappa_2\big(y_{xxx}+6y_x^2\big)+\kappa_1y_x-\tfrac{1}{4}x\right]
     \nonumber\\
   \qquad{} \times\big[4y_xy_{xxxxx}-4y_{xx}y_{xxxx}+2y_{xxx}^2+80y_x^2y_{xxx}+120y_x^4\nonumber\\
   \qquad {}
    +2\kappa_2\big(2y_xy_{xxx}-y_{xx}^2+8y_x^3\big)+2\kappa_1y_x^2-y\big]  =\tfrac{1}{16}(2\alpha-\epsilon)^2.\label{PII-O6-SD}
\end{gather}
The  B\"acklund transformation (\ref{PII-O6-BT1}),
(\ref{PII-O6-BT2}) and the equation (\ref{PII-O6-SD}) are not
given before.
\end{example}

\subsection{A hierarchy of a second-order fourth-degree equation}

In this subsection, we will generalize the B\"acklund
transformation given in \cite{AS2} between the second Painlev\'{e}
equation and a second-order fourth-degree equation.

Let
\begin{gather}\label{PII-BT2-1}
y=D_x^{-1}\Bigg[u_x\Bigg(\mathcal{R}_{_{\rm II}}^nu+\sum_{j=1}^{n-1}\kappa_{j}\mathcal{R}_{_{\rm II}}^{j}u\Bigg)\Bigg]
-\tfrac{1}{2}xu^2-\alpha u.
\end{gather}
Dif\/ferentiating (\ref{PII-BT2-1}) and using (\ref{PII-h}), we f\/ind
\begin{gather}\label{PII-BT2-2}
    u^2+2y_x=0.
\end{gather}
Equations (\ref{PII-BT2-1}) and  (\ref{PII-BT2-2}) def\/ine a
B\"acklund transformation between    the second Painlev\'{e}
hierarchy (\ref{PII-h}) and a new hierarchy of dif\/ferential
equations for $y$.

In order to obtain the new hierarchy, we will prove that
\begin{gather}\label{p2-1}
 D_x^{-1}\big(u_x\mathcal{R}_{_{\rm II}}^ju\big) =-D_x^{-1}\left(\frac{y_{xx}}{y_x}\mathcal{S}_{_{\rm II}}^jy_x\right), \end{gather}
where $\mathcal{S}_{_{\rm II}}$ is  the recursion operator
\[
 \mathcal{S}_{_{\rm II}}=D^2_x-\frac{y_{xx}}{y_x}D_x-\frac{y_{xxx}}{2y_x}
 +\frac{3y^2_{xx}}{4y^2_x}+8y_x-4y_xD^{-1}_x\frac{y_{xx}}{y_x}.
\]
First of all, we will use induction to prove that
\begin{gather}\label{p2-2}
\mathcal{R}_{_{\rm II}}^ju=-\frac{2}{u}\mathcal{S}_{_{\rm II}}^jy_x.
\end{gather}
 Using (\ref{PII-BT2-2}), we f\/ind
\begin{gather}\label{u-x}
    u_x=-\frac{y_{xx}}{u},
 \qquad u_{xx}=-\frac{1}{u}\left(y_{xxx}-\frac{y^2_{xx}}{2y_x}\right).
\end{gather}
Hence
 \begin{gather*}
\mathcal{R}_{_{\rm II}}u=u_{xx}-2u^3
       =-\frac{1}{u}\left(y_{xxx}-\frac{y^2_{xx}}{2y_x}+8y_x^2\right)
         =-\frac{2}{u}\mathcal{S}_{_{\rm II}}y_x.
\end{gather*}
Thus (\ref{p2-2}) is true for $j=1$.

 Assume it is true for $j=k$. Then
\begin{gather*}
  \mathcal{R}_{_{\rm II}}^{k+1} u=- 2\mathcal{R}_{_{\rm II}}\frac{1}{u}\mathcal{S}_{_{\rm II}}^k y_x
  =-\frac{2}{u}\left\{D^2_x-\frac{2u_x}{u}D_x
   -\frac{u_{xx}}{u}+\frac{2u_{x}^2}{u^2}
   -4u^2+4u^2D_x^{-1}\frac{u_x}{u}\right\}\mathcal{S}_{_{\rm II}}^ky_x.
\end{gather*}
Using (\ref{u-x}) to substitute $u_x$ and $u_{xx}$ and using
(\ref{PII-BT2-2}) to substitute $u^2$, we f\/ind the result.

As a second step,  we use (\ref{p2-2}) to f\/ind
\[
 D_x^{-1}\big(u_x\mathcal{R}_{_{\rm II}}^ku\big)=-2D_x^{-1}\left(\frac{u_x}{u}\mathcal{S}_{_{\rm II}}^ky_x\right) .
\]
Thus using   (\ref{u-x}) to substitute $u_x$ and using
(\ref{PII-BT2-2}) to substitute $u^2$ we f\/ind (\ref{p2-1}).

Therefore (\ref{PII-BT2-1}) implies
\begin{gather}\label{PII-BT2-3}
    \alpha u=-y+xy_x-
    D_x^{-1}\Bigg[\frac{y_{xx}}{y_x} \Bigg(\mathcal{S}_{_{\rm II}}^ny_x
    +\sum_{j=1}^{n-1}\kappa_{j}\mathcal{S}_{_{\rm II}}^{j}y_x\Bigg)\Bigg].
\end{gather}
If $\alpha\neq 0$, then substituting $u$ from (\ref{PII-BT2-3})
into (\ref{PII-BT2-2}) we obtain the following hierarchy of
dif\/ferential   equations for $y$
\begin{gather}\label{PII-SD2}
   \Bigg(D_x^{-1}\Bigg[\frac{y_{xx}}{y_x} \Bigg(\mathcal{S}_{_{\rm II}}^ny_x
    +\sum_{j=1}^{n-1}\kappa_{j}\mathcal{S}_{_{\rm II}}^{j}y_x\Bigg)\Bigg]-xy_x+y\Bigg)^2+ 2\alpha^2 y_x=0.
\end{gather}
If  $\alpha=0$, then $y$ satisf\/ies the hierarchy
\[
        D_x^{-1}\Bigg[\frac{y_{xx}}{y_x} \Bigg(\mathcal{S}_{_{\rm II}}^ny_x
    +\sum_{j=1}^{n-1}\kappa_{j}\mathcal{S}_{_{\rm II}}^{j}y_x\Bigg)\Bigg]-xy_x+y=0.
\]

The f\/irst member of the hierarchy (\ref{PII-SD2}) is a
fourth-degree equation, whereas the other members are
second-degree equations. Now we   give some examples.

\begin{example}[$n=1$]
In the present case,  (\ref{PII-BT2-1}) reads
\begin{gather}\label{PII-O2-BT2-1}
    2y=u_x^2-u^4-xu^2-2\alpha u.
\end{gather}
Eliminating $u$ between (\ref{PII-BT2-2}) and (\ref{PII-O2-BT2-1})
yields the following  second-order fourth-degree equation for $y$
 \begin{gather}\label{PII-SOFD}
    \big[y_{xx}^2+8y_x^3-4y_x(xy_x-y)\big]^2+32\alpha^2y_x^3=0.
\end{gather}
The change of variables $w=2y$ transform (\ref{PII-SOFD}) into the
following equation
\begin{gather}\label{PII-SOFD2}
    \big[w_{xx}^2+4w_x^3-4w_x(xw_x-w)\big]^2+16\alpha^2y_x^3=0.
\end{gather}
Equation (\ref{PII-SOFD2}) was derived before \cite{AS2}.
\end{example}

\begin{example}[$n=2$]
When $n=2$, (\ref{PII-BT2-1}) reads
\begin{gather}\label{PII-O4-BT2-1}
    2y=2u_xu_{xxx}-u_{xx}^2-10u^2u_x^2+2u^6-xu^2-2\alpha u+\kappa_1\big(u_x^2-u^4\big).
\end{gather}
Equations (\ref{PII-BT2-2}) and (\ref{PII-O4-BT2-1}) give a
B\"acklund transformation between the second member of PII
hierarchy (\ref{PII-O4}) and the following
 fourth-order second-degree equation for $y$
\begin{gather}
    \Bigg[y_{xx}y_{xxxx}-\frac{3y_{xx}^2}{2y_x}\left(y_{xxx}-\frac{y_{xx}^2}{2y_x}\right)
    -\frac{1}{2}\left(y_{xxx}-\frac{y_{xx}^2}{2y_x}\right)^2\nonumber\\
    \qquad{}
    +10y_xy_{xx}^2+16y_x^4-2y_x(xy_x-y)
    +\frac{1}{2}\kappa_1\big(y_{xx}^2+8y_x^3\big)\Bigg]^2+8\alpha^2y_x^3=0.\label{PII-FOSD}
\end{gather}
Equation (\ref{PII-FOSD}) was given before \cite{AS1}.
\end{example}

\begin{example}[$n=3$]
In this case, (\ref{PII-BT2-1})   read
\begin{gather}
    2y=2u_xu_{xxxxx}-2u_{xx}u_{xxxx}+u^2_{xxx}-28u^2u_xu_{xxx}
     +14u^2u_{xx}^2-56uu_x^2u_{xx}-21u_x^4\label{PII-O6-BT2-1}\\
\phantom{2y=}{} +70u^4u_x^2-5u^8
    +\kappa_2\big(2u_xu_{xxx}-u_{xx}^2-10u^2u_x^2+2u^6\big)
     +\kappa_1\big(u_x^2-u^4\big)-xu^2-2\alpha u,\nonumber
\end{gather}
 and (\ref{PII-SD2}) has the form
\begin{gather}
     \Bigg[2y_{xx}y_{xxxxxx}-\left(2y_{xxx}+\frac{3y_{xx}^2}{y_x}\right)\left(y_{xxxxx}+\frac{5y_{xx}y_{xxxx}}{y_x}\right)
        +\left(y_{xxxx}-\frac{3y_{xx}y_{xxx}}{2y_x}+\frac{3y_{xx}^3}{4y_x^2}\right)^2  \nonumber\\
        \qquad{}
    +\left(2y_{xxx}-\frac{y_{xx}^2}{y_x}\right)\left(\frac{2y_{xx}y_{xxxx}}{y_x}+\frac{3y_{xxx}^2}{2y_x}
                                      -\frac{9y_{xx}^2y_{xxx}}{2y_x^2}+\frac{15y_{xx}^2}{8y_x^3}
                                      -7y_{xx}^2-14y_xy_{xxx}  \right)   \nonumber\\
\qquad{}+\frac{15y_{xx}^2}{2y_x^2}\left(3y_{xxx}^2-\frac{5y_{xx}^2y_{xxx}}{y_x}+\frac{7y_{xx}^4}{4y_x^3}\right)
       +\frac{21y_{xx}^4}{2y_x}+280y_x^2y_{xx}^2-150y_x^5
   -4y_x(xy_x-y)     \nonumber\\
    \qquad{}+2\kappa_2\left[y_{xx}y_{xxxx}-\frac{3y_{xx}^2}{2y_x}\left(y_{xxx}-\frac{y_{xx}^2}{2y_x}\right)
    -\frac{1}{2}\left(y_{xxx}-\frac{y_{xx}^2}{2y_x}\right)^2
    +10y_xy_{xx}^2+16y_x^4 \right]    \nonumber\\
  \qquad{}  +\kappa_1\big(y_{xx}^2+8y_x^3\big)\Bigg]^2+32\alpha^2y_x^3=0.\label{PII-SOSD}
\end{gather}
The B\"acklund transformation between the third member of PII
hierarchy (\ref{PII-O6}) and the new equation (\ref{PII-SOSD}) is
given by
  (\ref{PII-BT2-2}) and (\ref{PII-O6-BT2-1}).
\end{example}

\subsection{Auto-B\"acklund transformations for PII hierarchy}

In this subsection, we will use the SD-I.d hierarchy
(\ref{SD-I.d-h}) to derive auto-B\"acklund transformations for PII
hierarchy (\ref{PII-h}).

Let $u$ be solution of (\ref{PII-h}) with parameter $\alpha$ and
let $\bar{u}$ be solution of (\ref{PII-h}) with parameter~$\bar{\alpha}.$ Since (\ref{SD-I.d-h}) is invariant under the
transformation $2\alpha-\epsilon=-2\bar{\alpha}+\epsilon$, a
solution $y$ of (\ref{SD-I.d-h}) corresponds to two solutions $u$
and $\bar{u}$ of (\ref{PII-h}). The relation between $y$ and $u$
is given by (\ref{PII-QE}) and the relation between $y$ and
$\bar{u}$ is given by
\begin{gather}
 \Bigg(-x+H^n[y_x]+\sum_{j=1}^{n-1}\kappa_jH^j[y_x]\Bigg)\bar{u}^2
    - (2\bar{\alpha}-\epsilon )\bar{u}               \nonumber\\
\qquad{}+ 2D_x^{-1}y_{x}\Bigg(H_x^n[y_x]+\sum_{j=1}^{n-1}\kappa_jH_x^j[y_x]\Bigg)-2y=0 .\label{PII-QE'}
 \end{gather}
Subtracting  (\ref{PII-QE}) from (\ref{PII-QE'}), we obtain
\begin{gather}\label{ABT-1}
 \Bigg(-x+H^n[y_x]+\sum_{j=1}^{n-1}\kappa_jH^j[y_x]\Bigg)\big(\bar{u}^2-u^2\big)-(2\bar{\alpha}-\epsilon)\bar{u}
 +(2\alpha-\epsilon)u=0.
\end{gather}
Using $2\alpha-\epsilon=-2\bar{\alpha}+\epsilon$ and dividing by
$\bar{u}+u$, (\ref{ABT-1}) yields
\[
\Bigg(-x+H^n[y_x]+\sum_{j=1}^{n-1}\kappa_jH^j[y_x]\Bigg)(\bar{u}-u)+(2\alpha-\epsilon)=0.
\]
Now using (\ref{PII-BT-2}) to substitute $y_x$, we obtain the
following two auto-B\"acklund transformations for PII hierarchy
(\ref{PII-h})
\begin{gather}
     \bar{\alpha}=-\alpha+\epsilon,\qquad \epsilon=\pm 1,        \nonumber\\
    \bar{u}=u- \frac{(2\alpha-\epsilon)}{\Big(-x+H^n[\frac{1}{2}(\epsilon u_x-u^2)]
    +\sum\limits_{j=1}^{n-1}\kappa_jH^j[\frac{1}{2}(\epsilon u_x-u^2)]\Big)} .\label{ABT}
\end{gather}
These auto-B\"acklund transformations and the discrete symmetry
$\bar{u}=-u$, $\bar{\alpha}=-\alpha$ can be used to derive the
auto-B\"acklund transformations  given in  \cite{Kudr4,  R4}.

The auto-B\"acklund transformations (\ref{ABT}) can be used to
obtain inf\/inite hierarchies of solutions of the PII hierarchy
(\ref{PII-h}). For example, starting by the solution $u=0$,
$\alpha=0$ of (\ref{PII-h}), the auto-B\"acklund transformations
(\ref{ABT}) yields the new solution
$ \bar{u}=-\frac{\epsilon}{x}$, $\bar{\alpha}=\epsilon $. Now
applying the auto-B\"acklund transformations (\ref{ABT}) with
$\epsilon=1$ to the solution
$ \bar{u}=\frac{1}{x}$, $\bar{\alpha}=-1$, we obtain the new
solution $\bar{\bar{u}} =\frac{-2(x^{3}-2\kappa_1)}{x(
x^{3}+4\kappa_1) }$,  $\bar{\bar{\alpha}}=2.$

\pdfbookmark[1]{References}{ref}
\LastPageEnding

\end{document}